\documentclass[aps,prb,showpacs,floats,twocolumn]{revtex4-1}

\usepackage{graphicx}
\usepackage{amsmath}
\usepackage{amssymb}

\newcommand{\bne}{\begin{equation*}}
\newcommand{\ede}{\end{equation*}}
\newcommand{\bea}{\begin{eqnarray*}}
\newcommand{\eea}{\end{eqnarray*}}
\newcommand{\bnen}{\begin{equation}}
\newcommand{\eden}{\end{equation}}
\newcommand{\bnsn}{\begin{subequations}}
\newcommand{\edsn}{\end{subequations}}
\newcommand{\bean}{\begin{eqnarray}}
\newcommand{\eean}{\end{eqnarray}}
\newcommand{\bna}{\begin{array}}
\newcommand{\eda}{\end{array}}

\newcommand{\st}[1]{\left|#1\right\rangle}
\newcommand{\f}[2]{\frac{#1}{#2}}
\newcommand{\us}{\uparrow}
\newcommand{\ds}{\downarrow}
\newcommand{\h}{\widetilde{H}}
\newcommand{\hc}{H_{\rm c}}
\newcommand{\s}{\sigma}
\newcommand{\ep}{\varepsilon}
\newcommand{\epm}{\varepsilon_{\rm m}}
\newcommand{\uc}{U_{\rm c}}
\newcommand{\muB}{\mu_{\rm B}}
\newcommand{\dBl}{b_{\rm l}}
\newcommand{\dBr}{b_{\rm r}}
\newcommand{\tl}{t_{\rm l}}
\newcommand{\tr}{t_{\rm r}}
\newcommand{\jl}{J_{\rm l}}
\newcommand{\jr}{J_{\rm r}}
\newcommand{\tj}[1]{T_{\varphi\,0#1}}


\begin{document}
\title{Hyperfine-induced dephasing in three-electron spin qubits}

\author{Csaba G. P\'eterfalvi}
\author{Guido Burkard}
\affiliation{Department of Physics,
University of Konstanz,
D-78464 Konstanz, Germany}


\begin{abstract}
We calculate the pure dephasing time of three-electron exchange-only qubits due to interaction with the nuclear hyperfine field. Within the $S=S_z=1/2$ spin subspace, we derive formulas for the dephasing time in the $(1,1,1)$ charge region and in the neighboring charge sectors coupled by tunneling. The nuclear field and the tunneling are taken into account in a second order approximation. The analytical solutions accurately reproduce the numerical evaluation of the full problem, and in comparison with existing experimental data, we find that the dephasing times are longer but on the same timescale as for single spins. Our analysis also applies to the resonant exchange (RX), always-on exchange-only (AEON) and hybrid qubits.
\end{abstract}

\maketitle

\section{Introduction}

Exchange-only qubits have attracted much attention due to their valuable feature of providing full control over the qubit by electrical gating of the dots themselves and the tunnel barriers in between. This can be seen as an evolution of qubit implementations in solid state systems that started with single-spin qubits,\cite{loss_quantum_1998} followed by singlet-triplet qubits\cite{taylor_fault-tolerant_2005} in double dots to arrive eventually to linearly arranged triple quantum dots that are controlled via tunnel couplings to the middle dot.\cite{divincenzo_universal_2000,medford_quantum-dot-based_2013,russ_asymmetric_2015,fei_characterizing_2015,shim_charge-noise-insensitive_2016,russ_three-electron_2016} All these systems are prone to decoherence on various time scales due to both magnetic and electrical noise. Electrical noise is always present due to fluctuations of the potential on the gates or background noise in the host material. This can be addressed by operating the qubit at the so-called ``sweet-spots''.\cite{russ_asymmetric_2015,fei_characterizing_2015,shim_charge-noise-insensitive_2016} Magnetic noise is also important in case nonzero nuclear spins are present in the vicinity of the qubit. This problem is particularly severe for example in GaAs in comparison to Si where the natural concentration of ${}^{29}$Si with nonzero spin is relatively low (around 5\%). But even in Si heterostructures, isotope purification is often the answer if further expansion of the dephasing time is needed. This underlines the necessity of studying decoherence and dephasing due to nuclear magnetic fields in exchange-only qubits,\cite{ladd_hyperfine-induced_2012,medford_quantum-dot-based_2013,mehl_noise_2013,hung_decoherence_2014} as well as developing dynamical methods to correct decohering qubits.\cite{hickman_dynamically_2013,zhang_benchmarking_2016,malinowski_symmetric_2017} This research has already progressed much for single-spin qubits,\cite{merkulov_electron_2002,khaetskii_electron_2002,coish_hyperfine_2004,deng_electron-spin_2008,cywinski_electron_2009,cywinski_pure_2009,coish_free-induction_2010,cywinski_dephasing_2011,assali_hyperfine_2011} and for singlet-triplet qubits as well.\cite{coish_singlet-triplet_2005,johnson_triplet-singlet_2005,petta_pulsed-gate_2005,petta_coherent_2005,bluhm_dephasing_2011,maune_coherent_2012,hung_hyperfine_2013,eng_isotopically_2015} Much of these results can also be found in a number of review articles.\cite{hanson_spins_2007,chirolli_decoherence_2008,russ_three-electron_2016}

In this paper, we try to further enrich our understanding of the role of hyperfine interaction in dephasing in exchange-only qubits. Within the $S=S_z=1/2$ spin subspace, which is decoherence-free against noise in a uniform magnetic field, we explore the $(1,1,1)$ charge sector and its surrounding, see in Fig.~\ref{fig:charge-stability} and~\ref{fig:lowest-two-states}. We derive analytic formulas for the dephasing time $T_\varphi$ with different logical qubit basis in the aforementioned charge sectors. We take the random nuclear field into account by averaging the density matrix over an ensemble of magnetic fields, thus obtaining a dephasing time (generally also denoted by $T_2^*$) which does not include the $T_1$ relaxation, and in this sense, characterizing the \emph{pure} dephasing of the qubit. We then evaluate and discuss our findings, their accuracy and symmetries, and compare them to results from the existing literature.

\section{Theoretical model}

\begin{figure}[t]
  \includegraphics[width=\columnwidth]{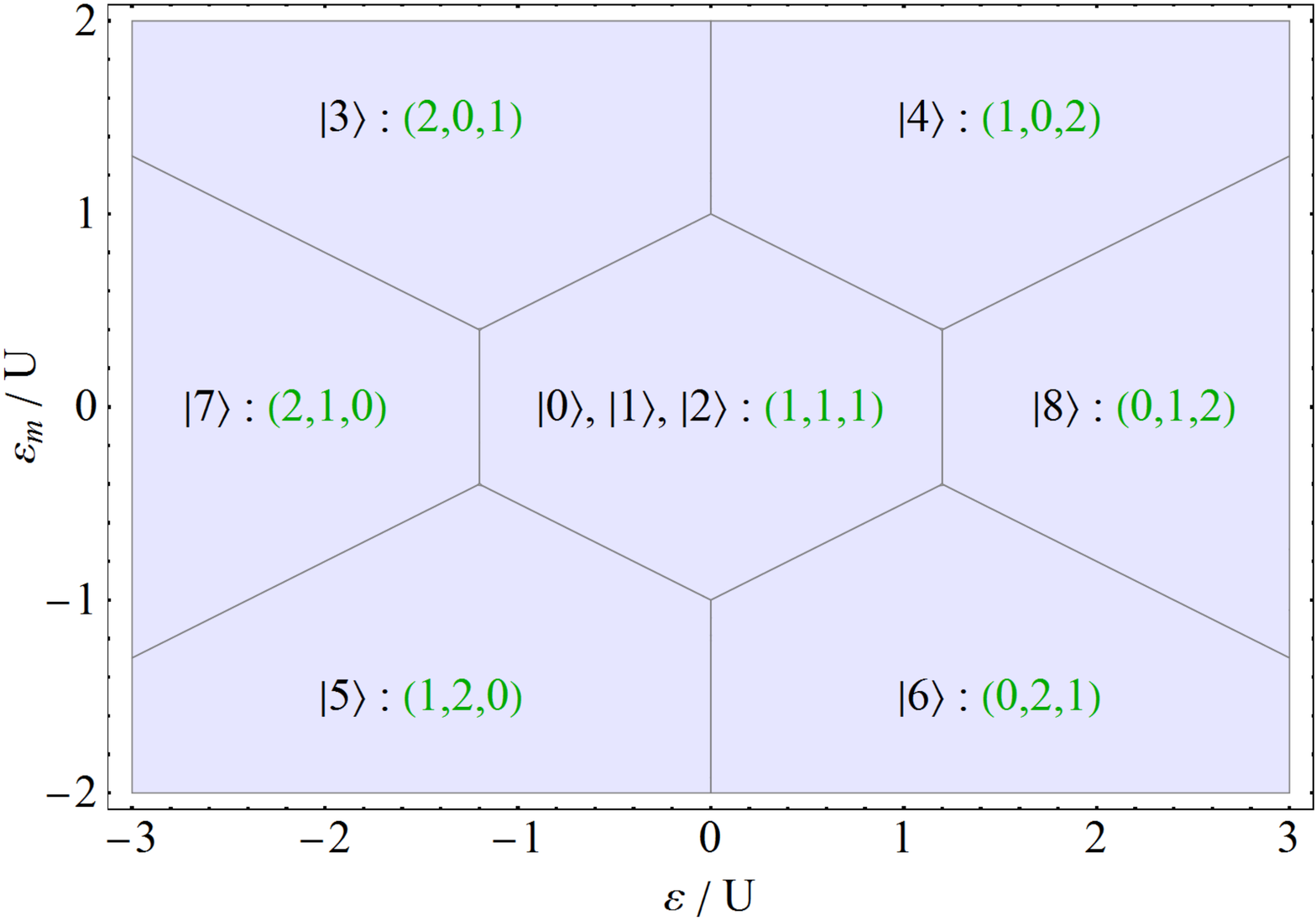}
	\caption{Charge stability diagram: lowest-energy charge states as functions of the detunings $\ep$ and $\epm$ with $\uc=0.2\,U$. In the absence of tunneling and hyperfine field, the states $\st0$, $\st1$ and $\st2$ are degenerate.\label{fig:charge-stability}}
\end{figure}

In our model, we consider a basis that consists of all three electron states with a total spin $S=1/2$ and a $z$-projection $S_z=1/2$:
\bnsn
\bean
\st0&=&\f{1}{\sqrt{2}}\left(\st{\us\us\ds}-\st{\ds\us\us}\right),\\
\st1&=&\f{1}{\sqrt{6}}\left(2\st{\us\ds\us}-\st{\us\us\ds}-\st{\ds\us\us}\right),\\
\st3&=&\f{1}{\sqrt{2}}\left(\st{\us\ds}_1-\st{\ds\us}_1\right)\st{\cdot}_2\st{\us}_3,\\
\st4&=&\f{1}{\sqrt{2}}\st{\us}_1\st{\cdot}_2\left(\st{\us\ds}_3-\st{\ds\us}_3\right),\\
\st5&=&\f{1}{\sqrt{2}}\st{\us}_1\left(\st{\us\ds}_2-\st{\ds\us}_2\right)\st{\cdot}_3,\\
\st6&=&\f{1}{\sqrt{2}}\st{\cdot}_1\left(\st{\us\ds}_2-\st{\ds\us}_2\right)\st{\us}_3,\\
\st7&=&\f{1}{\sqrt{2}}\left(\st{\us\ds}_1-\st{\ds\us}_1\right)\st{\us}_2\st{\cdot}_3,\\
\st8&=&\f{1}{\sqrt{2}}\st{\cdot}_1\st{\us}_2\left(\st{\us\ds}_3-\st{\ds\us}_3\right),
\eean\label{eq:basis}%
\edsn
where the subscript numbers the dot occupied by electron(s) with the given spin orientation, while $\st{\cdot}$ denotes an empty dot. We include an additional leakage state $\st2$ with a total spin of $S=3/2$ and $S_z=1/2$ because it is coupled to states $\st0$ and $\st1$ by the hyperfine field:
\addtocounter{equation}{-1}
\bnsn
\addtocounter{equation}{8}
\bean
\st2&=&\f{1}{\sqrt{3}}\left(\st{\ds\us\us}+\st{\us\ds\us}+\st{\us\us\ds}\right).
\eean
\edsn
States with different $z$-components can be split off with an external magnetic field. The states $\st0$, $\st1$ and $\st2$ belong to the charge state $(1,1,1)$, while in the other states, the electrons fill up the TQD according to the charge stability diagram, see Fig.~\ref{fig:charge-stability}.

The second quantized Hamiltonian of our model takes the form
\bean
&&H = \sum_i \ep_i n_i + U_0\sum_i n_{i\us} n_{i\ds} + \uc\sum_{\langle i,j \rangle} n_i n_j \\
&&+\!\!\!\sum_{{\langle i,j \rangle},\s} t_{i,j}\left(c_{i,\s}^\dagger c_{j,\s} + c_{j,\s}^\dagger c_{i,\s}\right) + \f{g \muB}{2}\sum_i \delta B_i \left(n_{i\us}-n_{i\ds}\right)\!,\nonumber
\eean
where $\ep_i$ is the electrostatic potential, $n_i=n_{i\us}+n_{i\ds}$ is the total number of electrons in dot $i$ and the number of electrons with a specific spin orientation respectively. $U_0$ is the on-site Coulomb interaction potential in case of double occupancy and $\uc$ is due to interaction of electrons in neighboring sites $\langle i,j \rangle$. The sums run over $i,j=1,2,3$. The tunnel coupling between these sites is denoted by $t_{i,j}$ and $c_{i,\s}^{(\dagger)}$ annihilates (creates) an electron in dot $i$ with spin $\s$. For simplicity, we neglect the external magnetic field, it serves only to split off states with different $S_z$ spin projections. In the Zeeman term, we only take into account the $z$-component of the Overhauser field $\delta B_i$ at dot $i$, where $g$ is the Land\'e $g$-factor and $\muB$ is the Bohr magneton.

\begin{figure}[b]
  \includegraphics[width=\columnwidth]{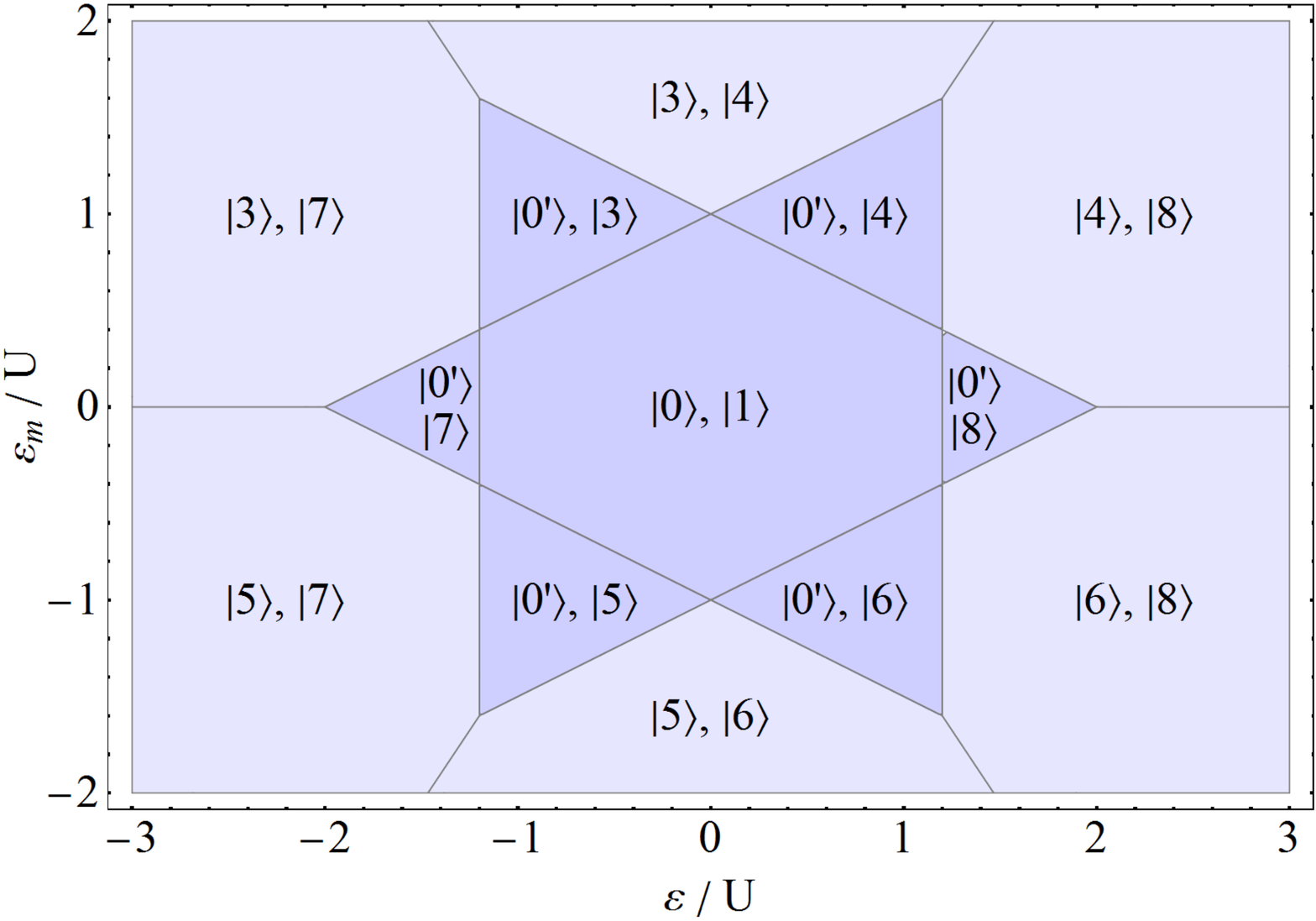}
	\caption{Partitioning of the $\ep$--$\epm$ space by the two lowest states with $\uc=0.2\,U$. Here $\st{0'}$ denotes the lower of the two states we obtain after the hybridization of $\st0$ and $\st1$. The regions of interest of this paper are shaded. \label{fig:lowest-two-states}}
\end{figure}

In the basis defined in~(\ref{eq:basis}), the matrix representation of the Hamiltonian takes the following form:
\bean
H&=&\left(\begin{array}{c|c}
 H_{01} & V \\ [0.3ex]
\cline{1-2}
\rule{0pt}{2.5ex}
 V^\dagger & \hc
\end{array}\right)\mbox{, where}\\
 H_{01}&=&\left(\begin{array}{ccc}
 0 & -\f{1}{\sqrt{3}}b_+ & \sqrt{\f23}b_+ \\
 -\f{1}{\sqrt{3}}b_+ & \f23 b_- & \f{\sqrt{2}}3 b_- \\
 \sqrt{\f23}b_+ & \f{\sqrt{2}}3 b_- & \f13 b_- 
\end{array}\right),\\
 V&=&\left(\begin{array}{cccccc}
 \f1{\sqrt{2}}\tl & \f1{\sqrt{2}}\tr & \f1{\sqrt{2}}\tr & \f1{\sqrt{2}}\tl & 0 & 0 \\
 \sqrt{\f32}\tl & -\sqrt{\f32}\tr & -\sqrt{\f32}\tr & \sqrt{\f32}\tl & 0 & 0 \\
 0 & 0 & 0 & 0 & 0 & 0
\end{array}\right),
\eean
and $\hc$ can be found in the Appendix~\ref{apx:Hc}.
The block $H_{01}$ describes the central $(1,1,1)$ region, where the energy scale is shifted so that these states are degenerate at 0 without tunneling and hyperfine interaction. The latter is characterized by $b_\pm=\dBl \pm \dBr$, where $\dBl=g\muB(\delta B_1-\delta B_2)/2$ and $\dBr=g\muB(\delta B_2-\delta B_3)/2$ denote the left and right hyperfine field gradients.
$V$ accounts for direct tunneling between the $(1,1,1)$ states and the states $\st{3}$, $\st{4}$, $\st{5}$ and $\st{6}$ with the tunnel coefficients $\tl=t_{1,2}$ and $\tr=t_{2,3}$. The states $\st{7}$ and $\st{8}$ are not coupled directly to the $(1,1,1)$ states, they are coupled in $\hc$ only to second order in $t_{\rm l,r}$. On the other hand, the state $\st2$ is not coupled to any other state by tunneling, but only by the nuclear field $b_\pm$.
The block $\hc$ describes the states around the $(1,1,1)$ region with a charge character. $\hc$ depends on $\ep=\ep_1-\ep_3$ measuring the detuning between the outer dots and on $\epm=\ep_2-(\ep_1+\ep_3)/2+\uc$, which is the relative detuning of the middle dot. It also depends on the Coulomb interaction parametrized by $\uc$ and $U=U_0-\uc$. $\uc$ is added to $\epm$ and subtracted from $U_0$ so that $H$ has a more symmetric form.
The lowest energy states of $H$ in the $\ep$--$\epm$ space can be seen in Fig.~\ref{fig:charge-stability}.
The lowest two states however (other than the leakage state) are the candidates for the logical qubit states, see Fig.~\ref{fig:lowest-two-states}.

To study dephasing in the $(1,1,1)$ regime and in its neighborhood where the hybridized $\st{0'}$ state is one of the lowest two states, we need to carry out separate calculations according to the partitioning in Fig.~\ref{fig:lowest-two-states}. In every region, an effective, reduced Hamiltonian needs to be found so that the problem is tractable analytically. We start with the central, $\st0-\st1$ region. We assume that inside this region, not too close to its borders, all the states other than the first three are far away in energy, and if the tunnel couplings are small enough, we can obtain an accurate approximation by applying the Schrieffer-Wolff transformation\cite{winkler_spin-orbit_2003} to $H$ keeping only $\st0$, $\st1$ and $\st2$ and transforming out the rest of the basis states. This involves the determination of a unitary operator $e^{-S}$ so that the basis transformation $\h=e^{-S}He^{S}$ leaves us with a block-diagonal $\h$ where the block with the states we are interested in is effectively decoupled from the rest of the states. We calculate this unitary operator with $S\propto t_{\rm l,r}$ in the absence of the hyperfine field, and then we apply it to the full Hamiltonian $H$ with finite hyperfine field. This is providing us with an effective Hamiltonian in a transformed basis $|\widetilde{n}\rangle=e^{-S}\st{n}$ with $n=0\dots8$. Up to second order in tunnel couplings, the resulting Hamiltonian for the $\st0-\st1$ region in the basis ${|\widetilde0\rangle,|\widetilde1\rangle,|\widetilde2\rangle}$ is
\bnen
\h_{01}=\left(\begin{array}{ccc}
 -\f14 J_+ & -\f1{\sqrt{3}} b_+ -\f{\sqrt{3}}{4}J_- & \sqrt{\f23} b_+ \\
 -\f1{\sqrt{3}} b_+ -\f{\sqrt{3}}{4}J_- & \f23 b_--\f34 J_+ & \f{\sqrt{2}}3 b_- \\
 \sqrt{\f23} b_+ & \f{\sqrt{2}}3 b_- & \f13 b_-
\end{array}\right)\!,
\eden
with exchange couplings defined as $J_\pm=\jl \pm \jr$, where
\bean
\jl&=&\f{4U\tl^2}{U^2-(\ep/2-\epm)^2}, \mbox{\ and}\\
\jr&=&\f{4U\tr^2}{U^2-(\ep/2+\epm)^2}.
\eean

\begin{figure}[t]
  \includegraphics[width=\columnwidth]{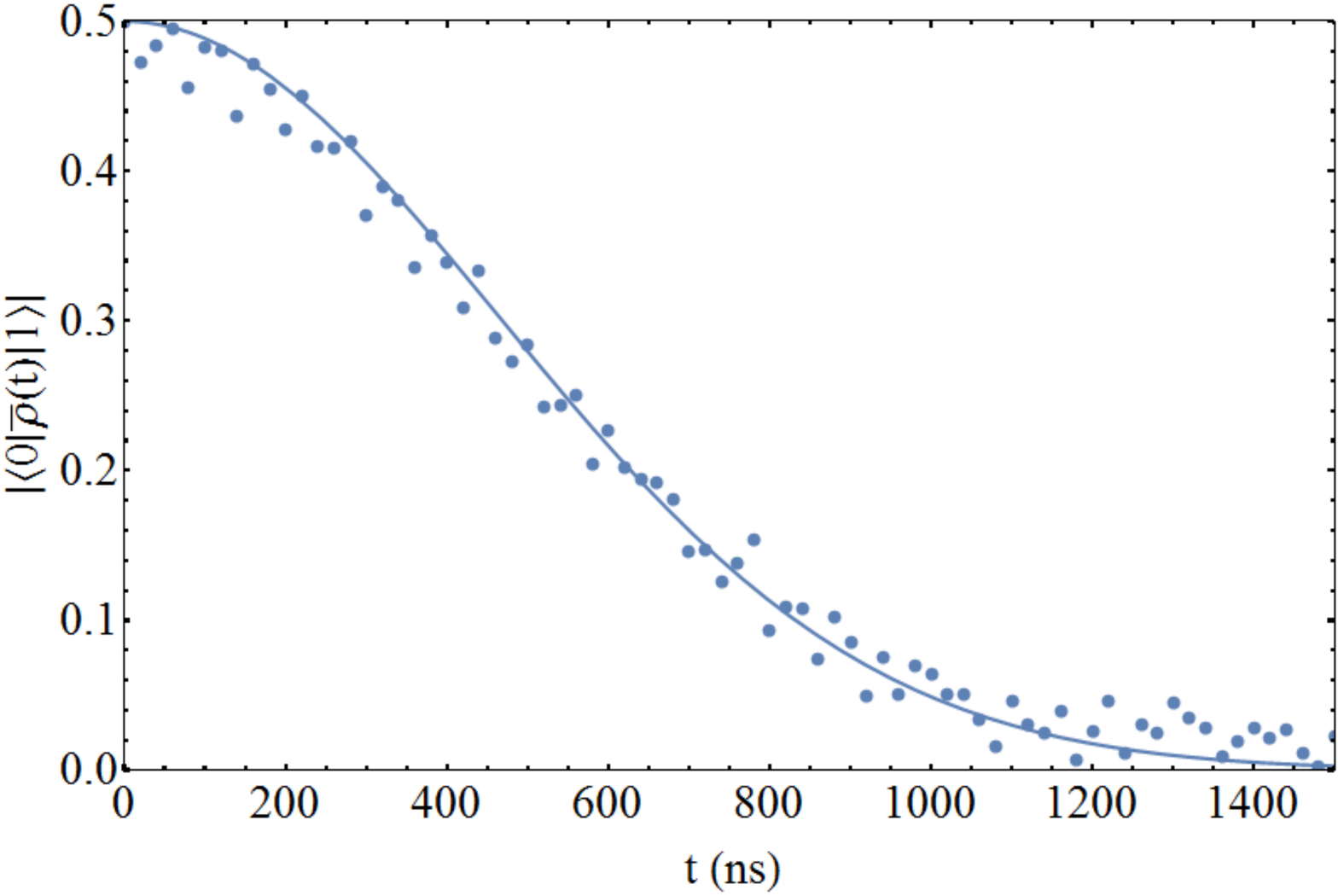}
	\caption{Temporal decay of the coherence $\langle0|\overline{\rho}(t)|1\rangle$. The solid line is obtained from Eq.~(\ref{eq:T01}), while the dots are calculated using the numerical evaluation of the full problem without approximations by taking the average of density matrices of states evolved by $H$ from the initial state of $(\st0+\st1)/\sqrt{2}$ with 1000 random realizations of the hyperfine field. Here $\ep/U=0.4$, $\epm/U=0.5$, $\uc/U=0.2$, $\tl/U=0.015$, $\tr/U=0.01$ and $g=2.0$, $\s_z=15\,\mu$T, which corresponds to natural Si.\cite{assali_hyperfine_2011} $\tj{1}=656\,$ns. \label{fig:rho12}}
\end{figure}

We next calculate the unitary operators that diagonalize $\h_{01}$ without the hyperfine field, then we apply this transformation to $\h_{01}$ together with the hyperfine field to obtain $\h_{01}'$. Here we assume that the dephasing is due to longitudinal noise, {\it i.e.} the wobbling of the energy levels of the Hamiltonian, while the transverse noise in $\h_{01}'$, which vanishes without the nuclear field, plays no role. For this reason, we ignore all the off-diagonal terms in $\h_{01}'$ and we solve a $2\times 2$ problem without a leakage state. As we shall see, numerical tests indeed justify this approximation. The time evolution operator $V_{01}(t)=\mathbb{I}-i\h_{01}'t/\hbar-\f12\h_{01}^{\prime 2}t^2/\hbar^2$ is constructed up to second order in time, where $\mathbb{I}$ is the identity operator and $\hbar$ is the reduced Planck constant. To extract the qubit dephasing time, $V_{01}(t)$ is applied to the initial state of $\f1{\sqrt{2}}(\st{0'}+\st{1'})$, where $\st{0'}$ and $\st{1'}$ are the two lowest states of $\h_{01}'$. The evolution of this pure state can be described by the corresponding density matrix $\rho_{01}(t)$, and its off-diagonal element $\langle0'|\rho_{01}(t)|1'\rangle$ characterizes the coherence of the state. The nuclear field however randomizes the matrix elements of $\rho_{01}(t)$, which we can take into account by averaging over $\delta B_1$, $\delta B_2$ and $\delta B_3$. Assuming that the $\delta B_i$ nuclear fields have an uncorrelated, normal distribution around 0 with a variance of $\s_z^2=\langle\delta B_i^2\rangle$ in the $z$-direction, we can calculate the ensemble averaged mixed state $\overline{\rho}_{01}(t)$ up to second order in the magnetic field. (The third powers also average to 0.) For the coherence term, in second order short-time approximation, we use the ansatz
\bnen
\langle0'|\overline{\rho}_{01}(t)|1'\rangle=\f12\left(1-c t^2\right)e^{it\omega-\f{t^2}{\tj{1}^2}}.
\eden
The dephasing time $\tj{1}$ appears in the exponent together with $\omega$, which is the energy difference between $\st{0'}$ and $\st{1'}$, while the constant $c$ is irrelevant for us at the moment. Using the power expansion of the coherence term, the dephasing time can be extracted as
\bnen
\tj{1}=\f{\sqrt{3}\hbar}{|g|\muB\s_z},\label{eq:T01}
\eden
in agreement with Hung~{\it et~al.}\cite{hung_decoherence_2014} It is interesting that in the $(1,1,1)$ region, the dephasing time depends neither on the detuning parameters $\ep$ and $\epm$, nor on the tunnel couplings $t_{\rm l,r}$. Nevertheless, this result agrees very well with the numerical solution of the full problem, see Fig.~\ref{fig:rho12}.

We now turn to the region in the charging diagram where the lowest two states are the $\st{0'}$ and $\st7$, see Fig.~\ref{fig:lowest-two-states}. In this case, we use the Schrieffer-Wolff transformation to separate the states $\st0$, $\st1$, $\st2$ and $\st7$ from the rest to obtain a reduced, $4\times 4$ Hamiltonian $\h_{07}$. Since state $\st7$ is not directly coupled to any of the first three states, the first $3\times3$ block of $\h_{07}$ is equal to $\h_{01}$. Similarly as before, we diagonalize $\h_{07}$ such that the off-diagonal elements of $\h_{07}'$ vanish for zero hyperfine fields. The extraction of the dephasing time leads us to
\bnen
\tj{7}=\f{2\sqrt{3}\hbar}{|g|\muB\s_z\sqrt{2+\f{\jl+\jr}{\sqrt{\jl^2-\jl\jr+\jr^2}}}}.\label{eq:T07}
\eden
As we see, unlike in the $(1,1,1)$ region, the dephasing time depends on the exchange couplings here. We can identify two limiting cases: if $\jl=\jr$ (for example $\tl=\tr$ and $\epm=0$), $\tj{7}=\tj{1}$, while if $\jl\gg\jr$ or $\jl\ll\jr$, $\tj{7}\approx 2\hbar/(|g|\muB\s_z)$. It can be shown that these are the two limiting cases for the minimum and the maximum of $\tj{7}$.

To obtain results for the opposite region with state $\st8$ being the lowest in energy, we need to interchange $\tl$ and $\tr$ and change the sign of $\ep$. This is effectively swapping $\jl$ and $\jr$, which leaves $\tj{7}$ unchanged, meaning that we can use the same formula in the opposite region,
\bnen
\tj{8}(\ep,\epm,\tl,\tr)=\tj{7}(\ep,\epm,\tl,\tr).\label{eq:T08}
\eden

In the case of the $\st{0'}-\st3$ region, we keep state $\st3$ together with the first three states, and we repeat the usual procedure to arrive to a somewhat more complex expression for the dephasing time, which can be found in Appendix~\ref{apx:Tj3}. The asymptotic expressions for two limiting cases however are the same simple expressions we have found before
\bean
\tj{3}(\ep,\epm,\tl\ll\tr) &\approx& \f{2\hbar}{|g|\muB\s_z}, \mbox{\ and}\\
\tj{3}(\ep,\epm,\tl\gg\tr) &\approx& \f{\sqrt{3}\hbar}{|g|\muB\s_z}.
\eean

Using again symmetry considerations, we can easily tell the dephasing time in the regions we have not covered yet:
\bean
\tj{4}(\ep,\epm,\tl,\tr)&=&\tj{3}(-\ep,\epm,\tr,\tl),\\
\tj{5}(\ep,\epm,\tl,\tr)&=&\tj{3}(\ep,-\epm,\tr,\tl), \mbox{\ and}\\
\tj{6}(\ep,\epm,\tl,\tr)&=&\tj{3}(-\ep,-\epm,\tl,\tr).\label{eq:T06}
\eean

It can be shown that in all seven regions, $T_{\varphi}$ only depends on the ratio of the tunnel couplings $\tl/\tr$ and not on their magnitude. This is consistent with the fact that the off-diagonal elements of the corresponding effective Hamiltonians $\h_{0n}'$ do not contribute to $\tj{n}$ in this approximation. The Schrieffer-Wolff transformation however relies on the assumption that the tunnel couplings are small relative to the smallest energy difference between the two sets of the basis states that are decoupled by this transformation. Here and also in general during the calculation, we neglected terms that were small in third or higher orders in the tunnel couplings. As a consequence, we should expect inaccuracies at the borders between the regions of interest, if we test our results within a distance from a border on the order of magnitude of the tunnel couplings.

\section{Discussion}

\begin{figure}[t]
  \includegraphics[width=\columnwidth]{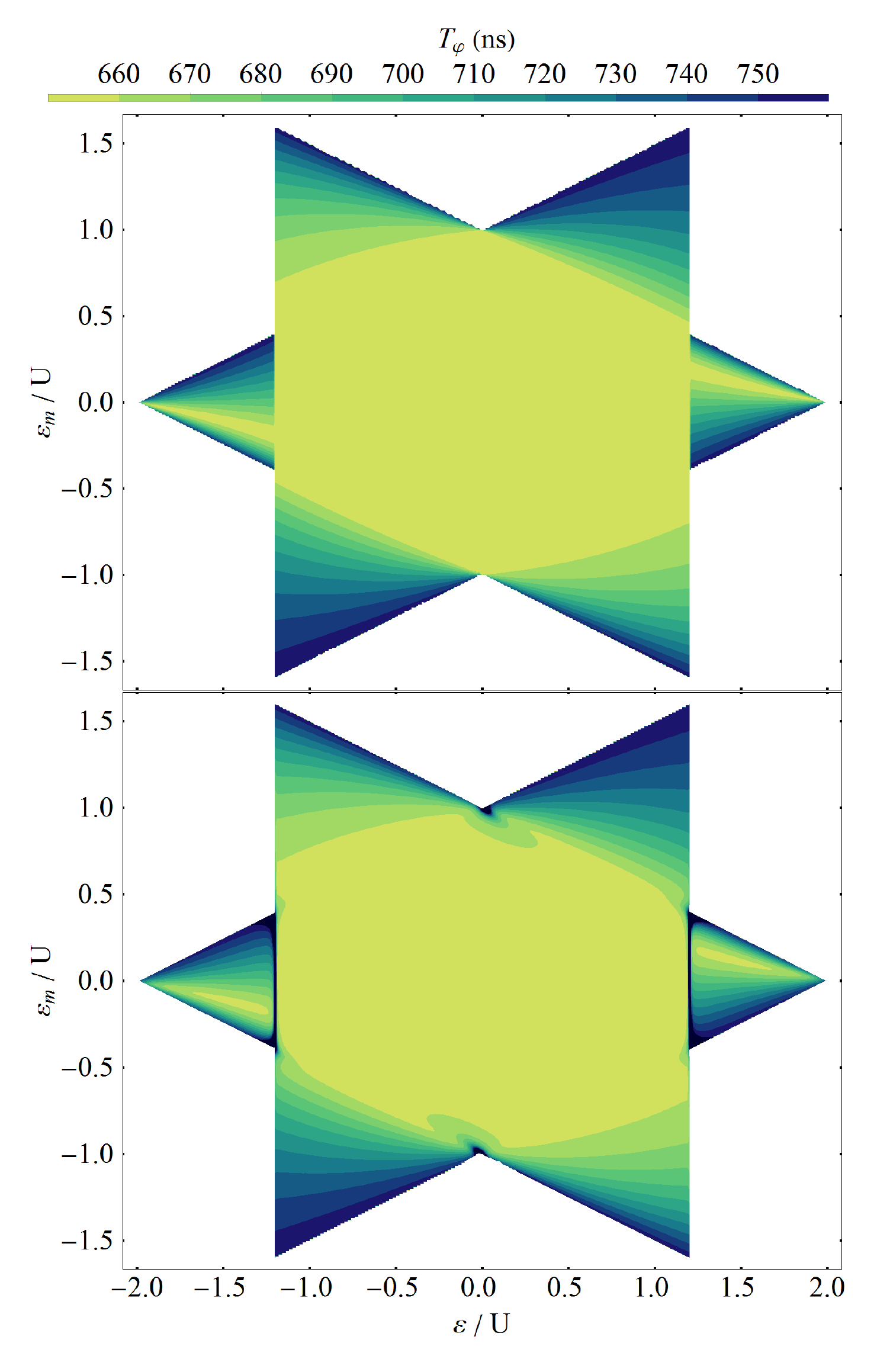}
	\caption{Dephasing time in nanoseconds for realistic input parameters: $\uc/U=0.2$, $\tl/U=0.015$, $\tr/U=0.01$ and $g=2.0$, $\s_z=15\,\mu$T, which corresponds to natural Si.\cite{assali_hyperfine_2011} In the upper figure, the analytic formulas for $\tj{n}$ are evaluated, while in the lower figure, the numerical solution of the full problem is plotted, which is also accurate close to the borders, see the main text.\label{fig:map-adiabatic}}
\end{figure}

We evaluated the formulas (\ref{eq:T01})-(\ref{eq:T06}) for a set of realistic parameters in the entire shaded region in Fig.~\ref{fig:lowest-two-states}. The result can be seen in Fig.~\ref{fig:map-adiabatic}. At the borders of the various regions, discontinuities may appear which are a consequence of using different basis states in the regions. These states typically hybridize close to the borders, which cannot be taken into account in our model due to the Schrieffer-Wolff transformation. The discontinuity is apparent at the border with regions on the left and right, which can be expected from the fact that states $\st7$ and $\st8$ are only indirectly coupled to $\st0$ and $\st1$, and due to the second order approximation, this coupling is lost entirely in the effective, reduced Hamiltonians $\h_{07}'$ and $\h_{08}'$.
Nevertheless, a comparison to a numerical analysis reveals that due to this loose coupling there is indeed a very sharp, step-like change at these borders, and the overall agreement between the analytical and numerical results is very good (see upper and lower part of Fig.~\ref{fig:map-adiabatic}). The dephasing time plotted in the lower part of Fig.~\ref{fig:map-adiabatic} was calculated by the numerical evaluation of the full Hamiltonian $H$, followed by its diagonalization and the time evolution of an initial state consisting of an equal superposition of the lowest two eigenstates. This pure state is then mixed by the hyperfine field and the dephasing time is extracted from the density matrix the same way as we did before. With this calculation, we do not need to discriminate between important and negligible states and the result will remain valid also close to the borders. The darkest shade of blue, which can be seen in the numerical results at the borders in question, is not present in the colorbar for the sake of easier comparison with the analytical results. There is a narrow peak here reaching up to $2.5~\mu$s at $\epm=0$, where we have a sharp avoided crossing between the states that are only indirectly coupled to each other.

If one crosses these borders in the $\ep-\epm$ space quickly enough, the transitions will be non-adiabatic and we arrive to a superposition of states not including the ground state in the given region. If at the beginning, we initialize the qubit in the lowest two states of the central region, then we will remain in this basis of $\h_{01}'$, and the dephasing time is constant and given by expression~(\ref{eq:T01}) for the whole region of interest.

If one moves slowly, however, and crosses the borders adiabatically keeping the qubit in the two lowest states, will create a hybrid qubit with a charge character, where two of the three electrons form a singlet state in one dot with zero spin, thus providing a natural protection against hyperfine noise. This explain why the dephasing time increases in the corresponding regions neighboring $(1,1,1)$.

There is an overall two-fold rotational symmetry in the plots meaning that $T_\varphi(\ep,\epm,\tl,\tr)=T_\varphi(-\ep,-\epm,\tl,\tr)$. The maxima can be found in the upper right and the lower left region if $\tl>\tr$ (see Fig.~\ref{fig:map-adiabatic}), and in the upper left and the lower right region if $\tl<\tr$. It can be shown that $T_\varphi(\ep,\epm,\tl,\tr)=T_\varphi(-\ep,\epm,\tr,\tl)$ is valid in general.

\begin{figure}[t]
  \includegraphics[width=\columnwidth]{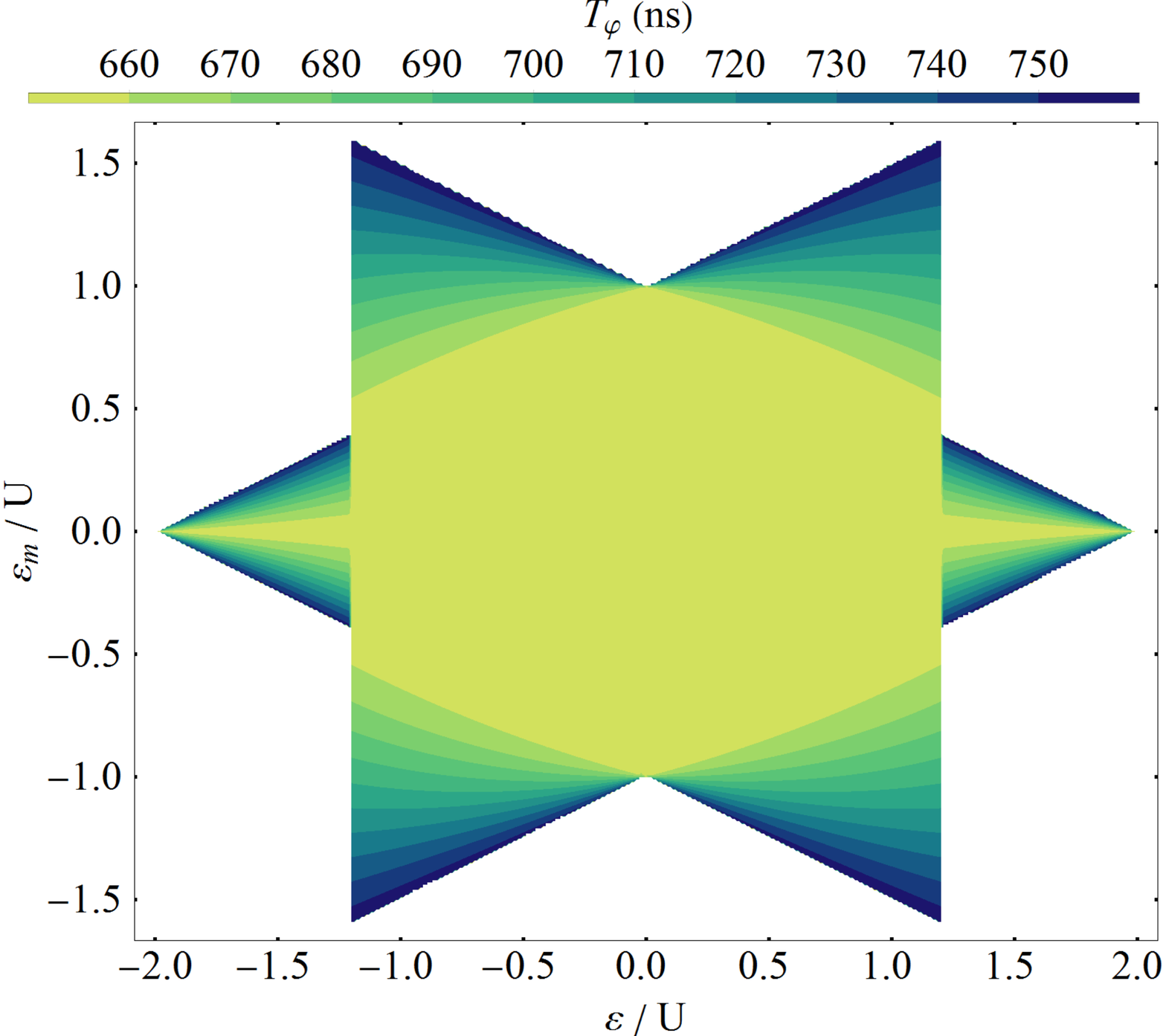}
	\caption{Dephasing time in nanoseconds for symmetric input parameters: $\uc/U=0.2$, $\tl/\tr=1$ and $g=2.0$, $\s=15\,\mu$T. \label{fig:map-adiabatic-symm}}
\end{figure}

The symmetry is even higher if $\tl=\tr$. In this case, we have two mirror planes and $T_\varphi(\ep,\epm,\tl,\tl)=T_\varphi(-\ep,\epm,\tl,\tl)=T_\varphi(\ep,-\epm,\tl,\tl)$, as shown in Fig.~\ref{fig:map-adiabatic-symm}.
It follows from (\ref{eq:T01}), (\ref{eq:T07}) and (\ref{eq:T08}) that if $\tl=\tr$, then $T_\varphi(\ep,\epm=0,\tl,\tl)=\sqrt{3}\hbar/(|g|\muB\s_z)$ in all three regions along the $\ep$-axis ($\epm=0$), which is the shortest dephasing time. For typical values of $\s_z$, this minimal $T_\varphi$ can be found in Table~\ref{tab:Tj(s)}.
\begin{table}[h]
 \caption{$\tj{1}=\sqrt{3}\hbar/(|g|\muB\s_z)$ evaluated at typical hyperfine field strengths.\cite{assali_hyperfine_2011}}
 \vspace{1mm}
 \centering
 \begin{tabular}{| c || c | c | c |} 
 \hline \rule{0pt}{2.5ex}
  host material & GaAs &\ Natural Si\ \ &\ 800 ppm ${}^{29}$Si\ \ \\ [0.5ex]
  \hline
	$\s_z$ &\ 2.1 mT\ \ & 15 $\mu$T & 1.9 $\mu$T \\ [1ex]
  \hline
	$g$ & $-0.44$ & 2.0 & 2.0 \\ [1ex]
  \hline
	\ $\tj{1}\propto \s_z^{-1}g^{-1}$\ \ & 22 ns & 0.66 $\mu$s & 5.2 $\mu$s \\ [1ex]
 \hline
 \end{tabular}
 \label{tab:Tj(s)}
\end{table}
It turns out that these dephasing times are on the same timescale as for single spins,\cite{merkulov_electron_2002,hanson_spins_2007,hung_decoherence_2014} but to be more precise, they tend to be approximately a factor of 2 larger. Indeed, in GaAs, the single-spin dephasing time is usually found to be around 10~ns.\cite{petta_coherent_2005,johnson_triplet-singlet_2005,hanson_spins_2007} For natural Si, Maune~{\it et~al.} measured a dephasing time of $0.36\,\mu$s,\cite{maune_coherent_2012} and for purified Si with 800~ppm residual ${}^{29}$Si content, Eng~{\it et~al.} obtained 2.3~$\mu$s.\cite{eng_isotopically_2015}

\section{Conclusions}

Mapping the dephasing time of exchange-only qubits over a wider region in the parameter space defined by the gate potentials opens up the possibility for better designs of pulse sequences where the important distinction has to be made between gate operations and keeping the qubit intact as long as possible between subsequent gate operations. Understandably, the storage of the qubit state should be in relative protection from potentially competing noise sources, optionally even outside of the central $(1,1,1)$ region. For this reason, it is imperative to know how much the qubit is affected by noise at different coordinates in this parameter space, and how the map of hyperfine-induced dephasing relates to the map of the ``sweet-spots'' of charge noise-induced dephasing.\cite{russ_asymmetric_2015} Motivated by this end, we calculated the pure dephasing time of three-electron exchange-only qubits due to interaction with the nuclear hyperfine field. Within the $S=S_z=1/2$ subspace, we derived formulas for the dephasing time in the $(1,1,1)$ charge region and in the neighboring charge sectors coupled by tunneling. The random nuclear field is taken into account by averaging the density matrix to an ensemble of magnetic fields up to second order. The tunnel couplings and the time of the initial states' evolution are also approximated up to second order. The analytical solutions accurately reproduce the numerical evaluation of the full problem. A comparison with existing experimental data finds that dephasing of single-spins is generally faster by a factor of 2 than dephasing of three-electron spin qubits. We demonstrated however that dephasing in our system can be further reduced by a factor of $2/\sqrt{3}$ by moving the qubit to neighboring hybrid qubit regions where the singlet state of two electrons provide additional protection from nuclear noise.

\phantom{.}
\acknowledgments

We are grateful for valuable discussions with Maximilian Russ. This work was supported by the DFG program SFB~767 and by ARO through Grant No. W911NF-15-1-0149.

\appendix
\section{The Hamiltonian block $\hc$}\label{apx:Hc}

The block of $H$ that describes the states around the $(1,1,1)$ region with a charge character
\begin{widetext}
\bnen
\hc=\left(\begin{array}{cccccc}
 U-\dBr+\f12\ep-\epm & 0 & 0 & 0 & \tr & 0 \\
 0 & U+\dBl-\f12\ep-\epm & 0 & 0 & 0 & \tl \\
 0 & 0 & U+\dBl+\f12\ep+\epm & 0 & -\tl & 0 \\
 0 & 0 & 0 & U-\dBr-\f12\ep+\epm & 0 & -\tr \\
 \tr & 0 & -\tl & 0 & U+\uc+\ep & 0 \\
 0 & \tl & 0 & -\tr & 0 & U+\uc-\ep
\end{array}\right).
\eden
\end{widetext}
Note that the states $\st{7}$ and $\st{8}$ are coupled to the $(1,1,1)$ states only in this block and only to second order in the tunnel couplings $t_{\rm l,r}$. So that the Schrieffer-Wolff transformation works, the diagonal elements of $\hc$ must be much larger in absolute value than $t_{\rm l,r}$.

\section{Analytical expression for $\tj{3}$}\label{apx:Tj3}

In the case of the $\st{0'}-\st3$ region, we obtain the dephasing time
\bean
&&\tj{3}=\f{2\hbar}{|g|\muB\s_z\sqrt{c_7^2-c_7c_8+c_8^2}},\ \mbox{where}\label{eq:T03}\\
&&c_{\rm lr}=\sqrt{\jl^2-\jl\jr+\jr^2},\nonumber\\
&&c_3=2-(\ep-2\epm)/U,\ c_4=\tl^2/U,\nonumber\\
&&c_5=12 \left(\jl-\jr\right)^2 c_3^4 c_{\rm lr}^2+\Big(c_3\jr\left(16c_4+c_3\left(c_3\jl+2c_{\rm lr}\right)\right)\nonumber\\
&&-\left(8c_4+c_3^2\jl/2\right)\left(8c_4+c_3(c_3\jl/2-c_{\rm lr})\right)-4c_3^2\jr^2\Big)^2\nonumber\\
&&c_6=c_3 c_{\rm lr} \left(8c_4+c_3(c_3\jl/2-2\jr)\right)/c_5,\nonumber\\
&&c_7=1-2 c_6 \left(8 c_4+c_3^2 \jl/2\right)^2+4 c_6 \left(c_5 c_6+2 c_3^2 c_{\rm lr}^2\right) \jr/c_{\rm lr},\nonumber\\
&&\mbox{and\ }c_8=2 c_6 \Big(64 c_4^2+8 c_3 c_4 \left(c_3 \jl-c_{\rm lr}-2 \jr\right)\nonumber\\
&&+c_3^2 \left(\jl \left(c_3^2 \jl/2-\left(4+c_3\right) c_{\rm lr}\right)/2-c_3 \jl \jr+4 \jr^2\right)\Big).\nonumber
\eean


\begin{thebibliography}{34}%
\makeatletter
\providecommand \@ifxundefined [1]{%
 \@ifx{#1\undefined}
}%
\providecommand \@ifnum [1]{%
 \ifnum #1\expandafter \@firstoftwo
 \else \expandafter \@secondoftwo
 \fi
}%
\providecommand \@ifx [1]{%
 \ifx #1\expandafter \@firstoftwo
 \else \expandafter \@secondoftwo
 \fi
}%
\providecommand \natexlab [1]{#1}%
\providecommand \enquote  [1]{``#1''}%
\providecommand \bibnamefont  [1]{#1}%
\providecommand \bibfnamefont [1]{#1}%
\providecommand \citenamefont [1]{#1}%
\providecommand \href@noop [0]{\@secondoftwo}%
\providecommand \href [0]{\begingroup \@sanitize@url \@href}%
\providecommand \@href[1]{\@@startlink{#1}\@@href}%
\providecommand \@@href[1]{\endgroup#1\@@endlink}%
\providecommand \@sanitize@url [0]{\catcode `\\12\catcode `\$12\catcode
  `\&12\catcode `\#12\catcode `\^12\catcode `\_12\catcode `\%12\relax}%
\providecommand \@@startlink[1]{}%
\providecommand \@@endlink[0]{}%
\providecommand \url  [0]{\begingroup\@sanitize@url \@url }%
\providecommand \@url [1]{\endgroup\@href {#1}{\urlprefix }}%
\providecommand \urlprefix  [0]{URL }%
\providecommand \Eprint [0]{\href }%
\providecommand \doibase [0]{http://dx.doi.org/}%
\providecommand \selectlanguage [0]{\@gobble}%
\providecommand \bibinfo  [0]{\@secondoftwo}%
\providecommand \bibfield  [0]{\@secondoftwo}%
\providecommand \translation [1]{[#1]}%
\providecommand \BibitemOpen [0]{}%
\providecommand \bibitemStop [0]{}%
\providecommand \bibitemNoStop [0]{.\EOS\space}%
\providecommand \EOS [0]{\spacefactor3000\relax}%
\providecommand \BibitemShut  [1]{\csname bibitem#1\endcsname}%
\let\auto@bib@innerbib\@empty
\bibitem [{\citenamefont {Loss}\ and\ \citenamefont
  {DiVincenzo}(1998)}]{loss_quantum_1998}%
  \BibitemOpen
  \bibfield  {author} {\bibinfo {author} {\bibfnamefont {D.}~\bibnamefont
  {Loss}}\ and\ \bibinfo {author} {\bibfnamefont {D.~P.}\ \bibnamefont
  {DiVincenzo}},\ }\href {\doibase 10.1103/PhysRevA.57.120} {\bibfield
  {journal} {\bibinfo  {journal} {Phys. Rev. A}\ }\textbf {\bibinfo {volume}
  {57}},\ \bibinfo {pages} {120} (\bibinfo {year} {1998})}\BibitemShut
  {NoStop}%
\bibitem [{\citenamefont {Taylor}\ \emph {et~al.}(2005)\citenamefont {Taylor},
  \citenamefont {Engel}, \citenamefont {D\"ur}, \citenamefont {Yacoby},
  \citenamefont {Marcus}, \citenamefont {Zoller},\ and\ \citenamefont
  {Lukin}}]{taylor_fault-tolerant_2005}%
  \BibitemOpen
  \bibfield  {author} {\bibinfo {author} {\bibfnamefont {J.~M.}\ \bibnamefont
  {Taylor}}, \bibinfo {author} {\bibfnamefont {H.-A.}\ \bibnamefont {Engel}},
  \bibinfo {author} {\bibfnamefont {W.}~\bibnamefont {D\"ur}}, \bibinfo {author}
  {\bibfnamefont {A.}~\bibnamefont {Yacoby}}, \bibinfo {author} {\bibfnamefont
  {C.~M.}\ \bibnamefont {Marcus}}, \bibinfo {author} {\bibfnamefont
  {P.}~\bibnamefont {Zoller}}, \ and\ \bibinfo {author} {\bibfnamefont {M.~D.}\
  \bibnamefont {Lukin}},\ }\href {\doibase 10.1038/nphys174} {\bibfield
  {journal} {\bibinfo  {journal} {Nat Phys}\ }\textbf {\bibinfo {volume} {1}},\
  \bibinfo {pages} {177} (\bibinfo {year} {2005})}\BibitemShut {NoStop}%
\bibitem [{\citenamefont {DiVincenzo}\ \emph {et~al.}(2000)\citenamefont
  {DiVincenzo}, \citenamefont {Bacon}, \citenamefont {Kempe}, \citenamefont
  {Burkard},\ and\ \citenamefont {Whaley}}]{divincenzo_universal_2000}%
  \BibitemOpen
  \bibfield  {author} {\bibinfo {author} {\bibfnamefont {D.~P.}\ \bibnamefont
  {DiVincenzo}}, \bibinfo {author} {\bibfnamefont {D.}~\bibnamefont {Bacon}},
  \bibinfo {author} {\bibfnamefont {J.}~\bibnamefont {Kempe}}, \bibinfo
  {author} {\bibfnamefont {G.}~\bibnamefont {Burkard}}, \ and\ \bibinfo
  {author} {\bibfnamefont {K.~B.}\ \bibnamefont {Whaley}},\ }\href {\doibase
  10.1038/35042541} {\bibfield  {journal} {\bibinfo  {journal} {Nature}\
  }\textbf {\bibinfo {volume} {408}},\ \bibinfo {pages} {339} (\bibinfo {year}
  {2000})}\BibitemShut {NoStop}%
\bibitem [{\citenamefont {Medford}\ \emph {et~al.}(2013)\citenamefont
  {Medford}, \citenamefont {Beil}, \citenamefont {Taylor}, \citenamefont
  {Rashba}, \citenamefont {Lu}, \citenamefont {Gossard},\ and\ \citenamefont
  {Marcus}}]{medford_quantum-dot-based_2013}%
  \BibitemOpen
  \bibfield  {author} {\bibinfo {author} {\bibfnamefont {J.}~\bibnamefont
  {Medford}}, \bibinfo {author} {\bibfnamefont {J.}~\bibnamefont {Beil}},
  \bibinfo {author} {\bibfnamefont {J.~M.}\ \bibnamefont {Taylor}}, \bibinfo
  {author} {\bibfnamefont {E.~I.}\ \bibnamefont {Rashba}}, \bibinfo {author}
  {\bibfnamefont {H.}~\bibnamefont {Lu}}, \bibinfo {author} {\bibfnamefont
  {A.~C.}\ \bibnamefont {Gossard}}, \ and\ \bibinfo {author} {\bibfnamefont
  {C.~M.}\ \bibnamefont {Marcus}},\ }\href {\doibase
  10.1103/PhysRevLett.111.050501} {\bibfield  {journal} {\bibinfo  {journal}
  {Phys. Rev. Lett.}\ }\textbf {\bibinfo {volume} {111}},\ \bibinfo {pages}
  {050501} (\bibinfo {year} {2013})}\BibitemShut {NoStop}%
\bibitem [{\citenamefont {Russ}\ and\ \citenamefont
  {Burkard}(2015)}]{russ_asymmetric_2015}%
  \BibitemOpen
  \bibfield  {author} {\bibinfo {author} {\bibfnamefont {M.}~\bibnamefont
  {Russ}}\ and\ \bibinfo {author} {\bibfnamefont {G.}~\bibnamefont {Burkard}},\
  }\href {\doibase 10.1103/PhysRevB.91.235411} {\bibfield  {journal} {\bibinfo
  {journal} {Phys. Rev. B}\ }\textbf {\bibinfo {volume} {91}},\ \bibinfo
  {pages} {235411} (\bibinfo {year} {2015})}\BibitemShut {NoStop}%
\bibitem [{\citenamefont {Fei}\ \emph {et~al.}(2015)\citenamefont {Fei},
  \citenamefont {Hung}, \citenamefont {Koh}, \citenamefont {Shim},
  \citenamefont {Coppersmith}, \citenamefont {Hu},\ and\ \citenamefont
  {Friesen}}]{fei_characterizing_2015}%
  \BibitemOpen
  \bibfield  {author} {\bibinfo {author} {\bibfnamefont {J.}~\bibnamefont
  {Fei}}, \bibinfo {author} {\bibfnamefont {J.-T.}\ \bibnamefont {Hung}},
  \bibinfo {author} {\bibfnamefont {T.~S.}\ \bibnamefont {Koh}}, \bibinfo
  {author} {\bibfnamefont {Y.-P.}\ \bibnamefont {Shim}}, \bibinfo {author}
  {\bibfnamefont {S.~N.}\ \bibnamefont {Coppersmith}}, \bibinfo {author}
  {\bibfnamefont {X.}~\bibnamefont {Hu}}, \ and\ \bibinfo {author}
  {\bibfnamefont {M.}~\bibnamefont {Friesen}},\ }\href {\doibase
  10.1103/PhysRevB.91.205434} {\bibfield  {journal} {\bibinfo  {journal} {Phys.
  Rev. B}\ }\textbf {\bibinfo {volume} {91}},\ \bibinfo {pages} {205434}
  (\bibinfo {year} {2015})}\BibitemShut {NoStop}%
\bibitem [{\citenamefont {Shim}\ and\ \citenamefont
  {Tahan}(2016)}]{shim_charge-noise-insensitive_2016}%
  \BibitemOpen
  \bibfield  {author} {\bibinfo {author} {\bibfnamefont {Y.-P.}\ \bibnamefont
  {Shim}}\ and\ \bibinfo {author} {\bibfnamefont {C.}~\bibnamefont {Tahan}},\
  }\href {\doibase 10.1103/PhysRevB.93.121410} {\bibfield  {journal} {\bibinfo
  {journal} {Phys. Rev. B}\ }\textbf {\bibinfo {volume} {93}},\ \bibinfo
  {pages} {121410} (\bibinfo {year} {2016})}\BibitemShut {NoStop}%
\bibitem [{\citenamefont {Russ}\ and\ \citenamefont
  {Burkard}(2016)}]{russ_three-electron_2016}%
  \BibitemOpen
  \bibfield  {author} {\bibinfo {author} {\bibfnamefont {M.}~\bibnamefont
  {Russ}}\ and\ \bibinfo {author} {\bibfnamefont {G.}~\bibnamefont {Burkard}},\
  }\href {http://arxiv.org/abs/1611.09106} {\bibfield  {journal} {\bibinfo
  {journal} {arXiv:1611.09106}}} \BibitemShut {NoStop}%
\bibitem [{\citenamefont {Ladd}(2012)}]{ladd_hyperfine-induced_2012}%
  \BibitemOpen
  \bibfield  {author} {\bibinfo {author} {\bibfnamefont {T.~D.}\ \bibnamefont
  {Ladd}},\ }\href {\doibase 10.1103/PhysRevB.86.125408} {\bibfield  {journal}
  {\bibinfo  {journal} {Phys. Rev. B}\ }\textbf {\bibinfo {volume} {86}},\
  \bibinfo {pages} {125408} (\bibinfo {year} {2012})}\BibitemShut {NoStop}%
\bibitem [{\citenamefont {Mehl}\ and\ \citenamefont
  {DiVincenzo}(2013)}]{mehl_noise_2013}%
  \BibitemOpen
  \bibfield  {author} {\bibinfo {author} {\bibfnamefont {S.}~\bibnamefont
  {Mehl}}\ and\ \bibinfo {author} {\bibfnamefont {D.~P.}\ \bibnamefont
  {DiVincenzo}},\ }\href {\doibase 10.1103/PhysRevB.87.195309} {\bibfield
  {journal} {\bibinfo  {journal} {Phys. Rev. B}\ }\textbf {\bibinfo {volume}
  {87}},\ \bibinfo {pages} {195309} (\bibinfo {year} {2013})}\BibitemShut
  {NoStop}%
\bibitem [{\citenamefont {Hung}\ \emph {et~al.}(2014)\citenamefont {Hung},
  \citenamefont {Fei}, \citenamefont {Friesen},\ and\ \citenamefont
  {Hu}}]{hung_decoherence_2014}%
  \BibitemOpen
  \bibfield  {author} {\bibinfo {author} {\bibfnamefont {J.-T.}\ \bibnamefont
  {Hung}}, \bibinfo {author} {\bibfnamefont {J.}~\bibnamefont {Fei}}, \bibinfo
  {author} {\bibfnamefont {M.}~\bibnamefont {Friesen}}, \ and\ \bibinfo
  {author} {\bibfnamefont {X.}~\bibnamefont {Hu}},\ }\href {\doibase
  10.1103/PhysRevB.90.045308} {\bibfield  {journal} {\bibinfo  {journal} {Phys.
  Rev. B}\ }\textbf {\bibinfo {volume} {90}},\ \bibinfo {pages} {045308}
  (\bibinfo {year} {2014})}\BibitemShut {NoStop}%
\bibitem [{\citenamefont {Hickman}\ \emph {et~al.}(2013)\citenamefont
  {Hickman}, \citenamefont {Wang}, \citenamefont {Kestner},\ and\ \citenamefont
  {Das~Sarma}}]{hickman_dynamically_2013}%
  \BibitemOpen
  \bibfield  {author} {\bibinfo {author} {\bibfnamefont {G.~T.}\ \bibnamefont
  {Hickman}}, \bibinfo {author} {\bibfnamefont {X.}~\bibnamefont {Wang}},
  \bibinfo {author} {\bibfnamefont {J.~P.}\ \bibnamefont {Kestner}}, \ and\
  \bibinfo {author} {\bibfnamefont {S.}~\bibnamefont {Das~Sarma}},\ }\href
  {\doibase 10.1103/PhysRevB.88.161303} {\bibfield  {journal} {\bibinfo
  {journal} {Phys. Rev. B}\ }\textbf {\bibinfo {volume} {88}},\ \bibinfo
  {pages} {161303} (\bibinfo {year} {2013})}\BibitemShut {NoStop}%
\bibitem [{\citenamefont {Zhang}\ \emph {et~al.}(2016)\citenamefont {Zhang},
  \citenamefont {Yang},\ and\ \citenamefont {Wang}}]{zhang_benchmarking_2016}%
  \BibitemOpen
  \bibfield  {author} {\bibinfo {author} {\bibfnamefont {C.}~\bibnamefont
  {Zhang}}, \bibinfo {author} {\bibfnamefont {X.-C.}\ \bibnamefont {Yang}}, \
  and\ \bibinfo {author} {\bibfnamefont {X.}~\bibnamefont {Wang}},\ }\href
  {\doibase 10.1103/PhysRevA.94.042323} {\bibfield  {journal} {\bibinfo
  {journal} {Phys. Rev. A}\ }\textbf {\bibinfo {volume} {94}},\ \bibinfo
  {pages} {042323} (\bibinfo {year} {2016})}\BibitemShut {NoStop}%
\bibitem [{\citenamefont {Malinowski}\ \emph {et~al.}(2017)\citenamefont
  {Malinowski}, \citenamefont {Martins}, \citenamefont {Nissen}, \citenamefont
  {Fallahi}, \citenamefont {Gardner}, \citenamefont {Manfra}, \citenamefont
  {Marcus},\ and\ \citenamefont {Kuemmeth}}]{malinowski_symmetric_2017}%
  \BibitemOpen
  \bibfield  {author} {\bibinfo {author} {\bibfnamefont {F.~K.}\ \bibnamefont
  {Malinowski}}, \bibinfo {author} {\bibfnamefont {F.}~\bibnamefont {Martins}},
  \bibinfo {author} {\bibfnamefont {P.~D.}\ \bibnamefont {Nissen}}, \bibinfo
  {author} {\bibfnamefont {S.}~\bibnamefont {Fallahi}}, \bibinfo {author}
  {\bibfnamefont {G.~C.}\ \bibnamefont {Gardner}}, \bibinfo {author}
  {\bibfnamefont {M.~J.}\ \bibnamefont {Manfra}}, \bibinfo {author}
  {\bibfnamefont {C.~M.}\ \bibnamefont {Marcus}}, \ and\ \bibinfo {author}
  {\bibfnamefont {F.}~\bibnamefont {Kuemmeth}},\ }\href
  {http://arxiv.org/abs/1704.01298} {\bibfield  {journal} {\bibinfo  {journal}
  {arXiv:1704.01298}} }\BibitemShut {NoStop}%
\bibitem [{\citenamefont {Merkulov}\ \emph {et~al.}(2002)\citenamefont
  {Merkulov}, \citenamefont {Efros},\ and\ \citenamefont
  {Rosen}}]{merkulov_electron_2002}%
  \BibitemOpen
  \bibfield  {author} {\bibinfo {author} {\bibfnamefont {I.~A.}\ \bibnamefont
  {Merkulov}}, \bibinfo {author} {\bibfnamefont {A.~L.}\ \bibnamefont {Efros}},
  \ and\ \bibinfo {author} {\bibfnamefont {M.}~\bibnamefont {Rosen}},\ }\href
  {\doibase 10.1103/PhysRevB.65.205309} {\bibfield  {journal} {\bibinfo
  {journal} {Phys. Rev. B}\ }\textbf {\bibinfo {volume} {65}},\ \bibinfo
  {pages} {205309} (\bibinfo {year} {2002})}\BibitemShut {NoStop}%
\bibitem [{\citenamefont {Khaetskii}\ \emph {et~al.}(2002)\citenamefont
  {Khaetskii}, \citenamefont {Loss},\ and\ \citenamefont
  {Glazman}}]{khaetskii_electron_2002}%
  \BibitemOpen
  \bibfield  {author} {\bibinfo {author} {\bibfnamefont {A.~V.}\ \bibnamefont
  {Khaetskii}}, \bibinfo {author} {\bibfnamefont {D.}~\bibnamefont {Loss}}, \
  and\ \bibinfo {author} {\bibfnamefont {L.}~\bibnamefont {Glazman}},\ }\href
  {\doibase 10.1103/PhysRevLett.88.186802} {\bibfield  {journal} {\bibinfo
  {journal} {Phys. Rev. Lett.}\ }\textbf {\bibinfo {volume} {88}},\ \bibinfo
  {pages} {186802} (\bibinfo {year} {2002})}\BibitemShut {NoStop}%
\bibitem [{\citenamefont {Coish}\ and\ \citenamefont
  {Loss}(2004)}]{coish_hyperfine_2004}%
  \BibitemOpen
  \bibfield  {author} {\bibinfo {author} {\bibfnamefont {W.~A.}\ \bibnamefont
  {Coish}}\ and\ \bibinfo {author} {\bibfnamefont {D.}~\bibnamefont {Loss}},\
  }\href {\doibase 10.1103/PhysRevB.70.195340} {\bibfield  {journal} {\bibinfo
  {journal} {Phys. Rev. B}\ }\textbf {\bibinfo {volume} {70}},\ \bibinfo
  {pages} {195340} (\bibinfo {year} {2004})}\BibitemShut {NoStop}%
\bibitem [{\citenamefont {Deng}\ and\ \citenamefont
  {Hu}(2008)}]{deng_electron-spin_2008}%
  \BibitemOpen
  \bibfield  {author} {\bibinfo {author} {\bibfnamefont {C.}~\bibnamefont
  {Deng}}\ and\ \bibinfo {author} {\bibfnamefont {X.}~\bibnamefont {Hu}},\
  }\href {\doibase 10.1103/PhysRevB.78.245301} {\bibfield  {journal} {\bibinfo
  {journal} {Phys. Rev. B}\ }\textbf {\bibinfo {volume} {78}},\ \bibinfo
  {pages} {245301} (\bibinfo {year} {2008})}\BibitemShut {NoStop}%
\bibitem [{\citenamefont {Cywi\'nski}\ \emph
  {et~al.}(2009{\natexlab{a}})\citenamefont {Cywi\'nski}, \citenamefont
  {Witzel},\ and\ \citenamefont {Das~Sarma}}]{cywinski_electron_2009}%
  \BibitemOpen
  \bibfield  {author} {\bibinfo {author} {\bibfnamefont {{\L}.}~\bibnamefont
  {Cywi\'nski}}, \bibinfo {author} {\bibfnamefont {W.~M.}\ \bibnamefont
  {Witzel}}, \ and\ \bibinfo {author} {\bibfnamefont {S.}~\bibnamefont
  {Das~Sarma}},\ }\href {\doibase 10.1103/PhysRevLett.102.057601} {\bibfield
  {journal} {\bibinfo  {journal} {Phys. Rev. Lett.}\ }\textbf {\bibinfo
  {volume} {102}},\ \bibinfo {pages} {057601} (\bibinfo {year}
  {2009}{\natexlab{a}})}\BibitemShut {NoStop}%
\bibitem [{\citenamefont {Cywi\'nski}\ \emph
  {et~al.}(2009{\natexlab{b}})\citenamefont {Cywi\'nski}, \citenamefont
  {Witzel},\ and\ \citenamefont {Das~Sarma}}]{cywinski_pure_2009}%
  \BibitemOpen
  \bibfield  {author} {\bibinfo {author} {\bibfnamefont {{\L}.}~\bibnamefont
  {Cywi\'nski}}, \bibinfo {author} {\bibfnamefont {W.~M.}\ \bibnamefont
  {Witzel}}, \ and\ \bibinfo {author} {\bibfnamefont {S.}~\bibnamefont
  {Das~Sarma}},\ }\href {\doibase 10.1103/PhysRevB.79.245314} {\bibfield
  {journal} {\bibinfo  {journal} {Phys. Rev. B}\ }\textbf {\bibinfo {volume}
  {79}},\ \bibinfo {pages} {245314} (\bibinfo {year}
  {2009}{\natexlab{b}})}\BibitemShut {NoStop}%
\bibitem [{\citenamefont {Coish}\ \emph {et~al.}(2010)\citenamefont {Coish},
  \citenamefont {Fischer},\ and\ \citenamefont
  {Loss}}]{coish_free-induction_2010}%
  \BibitemOpen
  \bibfield  {author} {\bibinfo {author} {\bibfnamefont {W.~A.}\ \bibnamefont
  {Coish}}, \bibinfo {author} {\bibfnamefont {J.}~\bibnamefont {Fischer}}, \
  and\ \bibinfo {author} {\bibfnamefont {D.}~\bibnamefont {Loss}},\ }\href
  {\doibase 10.1103/PhysRevB.81.165315} {\bibfield  {journal} {\bibinfo
  {journal} {Phys. Rev. B}\ }\textbf {\bibinfo {volume} {81}},\ \bibinfo
  {pages} {165315} (\bibinfo {year} {2010})}\BibitemShut {NoStop}%
\bibitem [{\citenamefont {Cywi\'nski}(2011)}]{cywinski_dephasing_2011}%
  \BibitemOpen
  \bibfield  {author} {\bibinfo {author} {\bibfnamefont {{\L}.}~\bibnamefont
  {Cywi\'nski}},\ }\href {\doibase 10.12693/APhysPolA.119.576} {\bibfield
  {journal} {\bibinfo  {journal} {Acta Physica Polonica A}\ }\textbf {\bibinfo
  {volume} {119}},\ \bibinfo {pages} {576} (\bibinfo {year} {2011})},\ \bibinfo
  {note} {arXiv: 1009.4466}\BibitemShut {NoStop}%
\bibitem [{\citenamefont {Assali}\ \emph {et~al.}(2011)\citenamefont {Assali},
  \citenamefont {Petrilli}, \citenamefont {Capaz}, \citenamefont {Koiller},
  \citenamefont {Hu},\ and\ \citenamefont {Das~Sarma}}]{assali_hyperfine_2011}%
  \BibitemOpen
  \bibfield  {author} {\bibinfo {author} {\bibfnamefont {L.~V.~C.}\
  \bibnamefont {Assali}}, \bibinfo {author} {\bibfnamefont {H.~M.}\
  \bibnamefont {Petrilli}}, \bibinfo {author} {\bibfnamefont {R.~B.}\
  \bibnamefont {Capaz}}, \bibinfo {author} {\bibfnamefont {B.}~\bibnamefont
  {Koiller}}, \bibinfo {author} {\bibfnamefont {X.}~\bibnamefont {Hu}}, \ and\
  \bibinfo {author} {\bibfnamefont {S.}~\bibnamefont {Das~Sarma}},\ }\href
  {\doibase 10.1103/PhysRevB.83.165301} {\bibfield  {journal} {\bibinfo
  {journal} {Phys. Rev. B}\ }\textbf {\bibinfo {volume} {83}},\ \bibinfo
  {pages} {165301} (\bibinfo {year} {2011})}\BibitemShut {NoStop}%
\bibitem [{\citenamefont {Coish}\ and\ \citenamefont
  {Loss}(2005)}]{coish_singlet-triplet_2005}%
  \BibitemOpen
  \bibfield  {author} {\bibinfo {author} {\bibfnamefont {W.~A.}\ \bibnamefont
  {Coish}}\ and\ \bibinfo {author} {\bibfnamefont {D.}~\bibnamefont {Loss}},\
  }\href {\doibase 10.1103/PhysRevB.72.125337} {\bibfield  {journal} {\bibinfo
  {journal} {Phys. Rev. B}\ }\textbf {\bibinfo {volume} {72}},\ \bibinfo
  {pages} {125337} (\bibinfo {year} {2005})}\BibitemShut {NoStop}%
\bibitem [{\citenamefont {Johnson}\ \emph {et~al.}(2005)\citenamefont
  {Johnson}, \citenamefont {Petta}, \citenamefont {Taylor}, \citenamefont
  {Yacoby}, \citenamefont {Lukin}, \citenamefont {Marcus}, \citenamefont
  {Hanson},\ and\ \citenamefont {Gossard}}]{johnson_triplet-singlet_2005}%
  \BibitemOpen
  \bibfield  {author} {\bibinfo {author} {\bibfnamefont {A.~C.}\ \bibnamefont
  {Johnson}}, \bibinfo {author} {\bibfnamefont {J.~R.}\ \bibnamefont {Petta}},
  \bibinfo {author} {\bibfnamefont {J.~M.}\ \bibnamefont {Taylor}}, \bibinfo
  {author} {\bibfnamefont {A.}~\bibnamefont {Yacoby}}, \bibinfo {author}
  {\bibfnamefont {M.~D.}\ \bibnamefont {Lukin}}, \bibinfo {author}
  {\bibfnamefont {C.~M.}\ \bibnamefont {Marcus}}, \bibinfo {author}
  {\bibfnamefont {M.~P.}\ \bibnamefont {Hanson}}, \ and\ \bibinfo {author}
  {\bibfnamefont {A.~C.}\ \bibnamefont {Gossard}},\ }\href {\doibase
  10.1038/nature03815} {\bibfield  {journal} {\bibinfo  {journal} {Nature}\
  }\textbf {\bibinfo {volume} {435}},\ \bibinfo {pages} {925} (\bibinfo {year}
  {2005})}\BibitemShut {NoStop}%
\bibitem [{\citenamefont {Petta}\ \emph
  {et~al.}(2005{\natexlab{a}})\citenamefont {Petta}, \citenamefont {Johnson},
  \citenamefont {Yacoby}, \citenamefont {Marcus}, \citenamefont {Hanson},\ and\
  \citenamefont {Gossard}}]{petta_pulsed-gate_2005}%
  \BibitemOpen
  \bibfield  {author} {\bibinfo {author} {\bibfnamefont {J.~R.}\ \bibnamefont
  {Petta}}, \bibinfo {author} {\bibfnamefont {A.~C.}\ \bibnamefont {Johnson}},
  \bibinfo {author} {\bibfnamefont {A.}~\bibnamefont {Yacoby}}, \bibinfo
  {author} {\bibfnamefont {C.~M.}\ \bibnamefont {Marcus}}, \bibinfo {author}
  {\bibfnamefont {M.~P.}\ \bibnamefont {Hanson}}, \ and\ \bibinfo {author}
  {\bibfnamefont {A.~C.}\ \bibnamefont {Gossard}},\ }\href {\doibase
  10.1103/PhysRevB.72.161301} {\bibfield  {journal} {\bibinfo  {journal} {Phys.
  Rev. B}\ }\textbf {\bibinfo {volume} {72}},\ \bibinfo {pages} {161301}
  (\bibinfo {year} {2005}{\natexlab{a}})}\BibitemShut {NoStop}%
\bibitem [{\citenamefont {Petta}\ \emph
  {et~al.}(2005{\natexlab{b}})\citenamefont {Petta}, \citenamefont {Johnson},
  \citenamefont {Taylor}, \citenamefont {Laird}, \citenamefont {Yacoby},
  \citenamefont {Lukin}, \citenamefont {Marcus}, \citenamefont {Hanson},\ and\
  \citenamefont {Gossard}}]{petta_coherent_2005}%
  \BibitemOpen
  \bibfield  {author} {\bibinfo {author} {\bibfnamefont {J.~R.}\ \bibnamefont
  {Petta}}, \bibinfo {author} {\bibfnamefont {A.~C.}\ \bibnamefont {Johnson}},
  \bibinfo {author} {\bibfnamefont {J.~M.}\ \bibnamefont {Taylor}}, \bibinfo
  {author} {\bibfnamefont {E.~A.}\ \bibnamefont {Laird}}, \bibinfo {author}
  {\bibfnamefont {A.}~\bibnamefont {Yacoby}}, \bibinfo {author} {\bibfnamefont
  {M.~D.}\ \bibnamefont {Lukin}}, \bibinfo {author} {\bibfnamefont {C.~M.}\
  \bibnamefont {Marcus}}, \bibinfo {author} {\bibfnamefont {M.~P.}\
  \bibnamefont {Hanson}}, \ and\ \bibinfo {author} {\bibfnamefont {A.~C.}\
  \bibnamefont {Gossard}},\ }\href {\doibase 10.1126/science.1116955}
  {\bibfield  {journal} {\bibinfo  {journal} {Science}\ }\textbf {\bibinfo
  {volume} {309}},\ \bibinfo {pages} {2180} (\bibinfo {year}
  {2005}{\natexlab{b}})}\BibitemShut {NoStop}%
\bibitem [{\citenamefont {Bluhm}\ \emph {et~al.}(2011)\citenamefont {Bluhm},
  \citenamefont {Foletti}, \citenamefont {Neder}, \citenamefont {Rudner},
  \citenamefont {Mahalu}, \citenamefont {Umansky},\ and\ \citenamefont
  {Yacoby}}]{bluhm_dephasing_2011}%
  \BibitemOpen
  \bibfield  {author} {\bibinfo {author} {\bibfnamefont {H.}~\bibnamefont
  {Bluhm}}, \bibinfo {author} {\bibfnamefont {S.}~\bibnamefont {Foletti}},
  \bibinfo {author} {\bibfnamefont {I.}~\bibnamefont {Neder}}, \bibinfo
  {author} {\bibfnamefont {M.}~\bibnamefont {Rudner}}, \bibinfo {author}
  {\bibfnamefont {D.}~\bibnamefont {Mahalu}}, \bibinfo {author} {\bibfnamefont
  {V.}~\bibnamefont {Umansky}}, \ and\ \bibinfo {author} {\bibfnamefont
  {A.}~\bibnamefont {Yacoby}},\ }\href {\doibase 10.1038/nphys1856} {\bibfield
  {journal} {\bibinfo  {journal} {Nat Phys}\ }\textbf {\bibinfo {volume} {7}},\
  \bibinfo {pages} {109} (\bibinfo {year} {2011})}\BibitemShut {NoStop}%
\bibitem [{\citenamefont {Maune}\ \emph {et~al.}(2012)\citenamefont {Maune},
  \citenamefont {Borselli}, \citenamefont {Huang}, \citenamefont {Ladd},
  \citenamefont {Deelman}, \citenamefont {Holabird}, \citenamefont {Kiselev},
  \citenamefont {Alvarado-Rodriguez}, \citenamefont {Ross}, \citenamefont
  {Schmitz}, \citenamefont {Sokolich}, \citenamefont {Watson}, \citenamefont
  {Gyure},\ and\ \citenamefont {Hunter}}]{maune_coherent_2012}%
  \BibitemOpen
  \bibfield  {author} {\bibinfo {author} {\bibfnamefont {B.~M.}\ \bibnamefont
  {Maune}}, \bibinfo {author} {\bibfnamefont {M.~G.}\ \bibnamefont {Borselli}},
  \bibinfo {author} {\bibfnamefont {B.}~\bibnamefont {Huang}}, \bibinfo
  {author} {\bibfnamefont {T.~D.}\ \bibnamefont {Ladd}}, \bibinfo {author}
  {\bibfnamefont {P.~W.}\ \bibnamefont {Deelman}}, \bibinfo {author}
  {\bibfnamefont {K.~S.}\ \bibnamefont {Holabird}}, \bibinfo {author}
  {\bibfnamefont {A.~A.}\ \bibnamefont {Kiselev}}, \bibinfo {author}
  {\bibfnamefont {I.}~\bibnamefont {Alvarado-Rodriguez}}, \bibinfo {author}
  {\bibfnamefont {R.~S.}\ \bibnamefont {Ross}}, \bibinfo {author}
  {\bibfnamefont {A.~E.}\ \bibnamefont {Schmitz}}, \bibinfo {author}
  {\bibfnamefont {M.}~\bibnamefont {Sokolich}}, \bibinfo {author}
  {\bibfnamefont {C.~A.}\ \bibnamefont {Watson}}, \bibinfo {author}
  {\bibfnamefont {M.~F.}\ \bibnamefont {Gyure}}, \ and\ \bibinfo {author}
  {\bibfnamefont {A.~T.}\ \bibnamefont {Hunter}},\ }\href {\doibase
  10.1038/nature10707} {\bibfield  {journal} {\bibinfo  {journal} {Nature}\
  }\textbf {\bibinfo {volume} {481}},\ \bibinfo {pages} {344} (\bibinfo {year}
  {2012})}\BibitemShut {NoStop}%
\bibitem [{\citenamefont {Hung}\ \emph {et~al.}(2013)\citenamefont {Hung},
  \citenamefont {Cywi\'nski}, \citenamefont {Hu},\ and\ \citenamefont
  {Das~Sarma}}]{hung_hyperfine_2013}%
  \BibitemOpen
  \bibfield  {author} {\bibinfo {author} {\bibfnamefont {J.-T.}\ \bibnamefont
  {Hung}}, \bibinfo {author} {\bibfnamefont {{\L}.}~\bibnamefont {Cywi\'nski}},
  \bibinfo {author} {\bibfnamefont {X.}~\bibnamefont {Hu}}, \ and\ \bibinfo
  {author} {\bibfnamefont {S.}~\bibnamefont {Das~Sarma}},\ }\href {\doibase
  10.1103/PhysRevB.88.085314} {\bibfield  {journal} {\bibinfo  {journal} {Phys.
  Rev. B}\ }\textbf {\bibinfo {volume} {88}},\ \bibinfo {pages} {085314}
  (\bibinfo {year} {2013})}\BibitemShut {NoStop}%
\bibitem [{\citenamefont {Eng}\ \emph {et~al.}(2015)\citenamefont {Eng},
  \citenamefont {Ladd}, \citenamefont {Smith}, \citenamefont {Borselli},
  \citenamefont {Kiselev}, \citenamefont {Fong}, \citenamefont {Holabird},
  \citenamefont {Hazard}, \citenamefont {Huang}, \citenamefont {Deelman},
  \citenamefont {Milosavljevic}, \citenamefont {Schmitz}, \citenamefont {Ross},
  \citenamefont {Gyure},\ and\ \citenamefont {Hunter}}]{eng_isotopically_2015}%
  \BibitemOpen
  \bibfield  {author} {\bibinfo {author} {\bibfnamefont {K.}~\bibnamefont
  {Eng}}, \bibinfo {author} {\bibfnamefont {T.~D.}\ \bibnamefont {Ladd}},
  \bibinfo {author} {\bibfnamefont {A.}~\bibnamefont {Smith}}, \bibinfo
  {author} {\bibfnamefont {M.~G.}\ \bibnamefont {Borselli}}, \bibinfo {author}
  {\bibfnamefont {A.~A.}\ \bibnamefont {Kiselev}}, \bibinfo {author}
  {\bibfnamefont {B.~H.}\ \bibnamefont {Fong}}, \bibinfo {author}
  {\bibfnamefont {K.~S.}\ \bibnamefont {Holabird}}, \bibinfo {author}
  {\bibfnamefont {T.~M.}\ \bibnamefont {Hazard}}, \bibinfo {author}
  {\bibfnamefont {B.}~\bibnamefont {Huang}}, \bibinfo {author} {\bibfnamefont
  {P.~W.}\ \bibnamefont {Deelman}}, \bibinfo {author} {\bibfnamefont
  {I.}~\bibnamefont {Milosavljevic}}, \bibinfo {author} {\bibfnamefont {A.~E.}\
  \bibnamefont {Schmitz}}, \bibinfo {author} {\bibfnamefont {R.~S.}\
  \bibnamefont {Ross}}, \bibinfo {author} {\bibfnamefont {M.~F.}\ \bibnamefont
  {Gyure}}, \ and\ \bibinfo {author} {\bibfnamefont {A.~T.}\ \bibnamefont
  {Hunter}},\ }\href {\doibase 10.1126/sciadv.1500214} {\bibfield  {journal}
  {\bibinfo  {journal} {Sci Adv}\ }\textbf {\bibinfo {volume} {1}} (\bibinfo
  {year} {2015}),\ 10.1126/sciadv.1500214}\BibitemShut {NoStop}%
\bibitem [{\citenamefont {Hanson}\ \emph {et~al.}(2007)\citenamefont {Hanson},
  \citenamefont {Kouwenhoven}, \citenamefont {Petta}, \citenamefont {Tarucha},\
  and\ \citenamefont {Vandersypen}}]{hanson_spins_2007}%
  \BibitemOpen
  \bibfield  {author} {\bibinfo {author} {\bibfnamefont {R.}~\bibnamefont
  {Hanson}}, \bibinfo {author} {\bibfnamefont {L.~P.}\ \bibnamefont
  {Kouwenhoven}}, \bibinfo {author} {\bibfnamefont {J.~R.}\ \bibnamefont
  {Petta}}, \bibinfo {author} {\bibfnamefont {S.}~\bibnamefont {Tarucha}}, \
  and\ \bibinfo {author} {\bibfnamefont {L.~M.~K.}\ \bibnamefont
  {Vandersypen}},\ }\href {\doibase 10.1103/RevModPhys.79.1217} {\bibfield
  {journal} {\bibinfo  {journal} {Rev. Mod. Phys.}\ }\textbf {\bibinfo {volume}
  {79}},\ \bibinfo {pages} {1217} (\bibinfo {year} {2007})}\BibitemShut
  {NoStop}%
\bibitem [{\citenamefont {Chirolli}\ and\ \citenamefont
  {Burkard}(2008)}]{chirolli_decoherence_2008}%
  \BibitemOpen
  \bibfield  {author} {\bibinfo {author} {\bibfnamefont {L.}~\bibnamefont
  {Chirolli}}\ and\ \bibinfo {author} {\bibfnamefont {G.}~\bibnamefont
  {Burkard}},\ }\href {\doibase 10.1080/00018730802218067} {\bibfield
  {journal} {\bibinfo  {journal} {Advances in Physics}\ }\textbf {\bibinfo
  {volume} {57}},\ \bibinfo {pages} {225} (\bibinfo {year} {2008})}\BibitemShut
  {NoStop}%
\bibitem [{\citenamefont {Winkler}(2003)}]{winkler_spin-orbit_2003}%
  \BibitemOpen
  \bibfield  {author} {\bibinfo {author} {\bibfnamefont {R.}~\bibnamefont
  {Winkler}},\ }\href@noop {} {\emph {\bibinfo {title} {Spin-orbit {Coupling}
  {Effects} in {Two}-{Dimensional} {Electron} and {Hole} {Systems}}}}\
  (\bibinfo  {publisher} {Springer},\ \bibinfo {year}
  {2003})\BibitemShut {NoStop}%
\end{thebibliography}
%

\end{document}